\documentclass[a4paper,preprintnumbers,floatfix,superscriptaddress,pra,showpacs,notitlepage,longbibliography]{revtex4-1}
\usepackage[utf8]{inputenc}
\usepackage[T1]{fontenc}
\usepackage[sc,osf]{mathpazo}
\usepackage{mathtools}
\usepackage{amsfonts}
\usepackage{bbm}
\usepackage{booktabs}
\usepackage{diagbox}
\usepackage{float}
\usepackage[table]{xcolor}

\begin{document}

\title{Statistical mechanical approach of complex networks with weighted links}
\author{Rute Oliveira}
\affiliation{Federal University of Rio Grande do Norte, Departamento de F\'isica Te\'orica e Experimental, Natal-RN, 59078-900, Brazil.}
\author{Samuraí Brito}
\affiliation{Emerging Technology Group at Itau-Unibanco}
\affiliation{International Institute of Physics, Federal University of Rio Grande do Norte, 59070-405 Natal, Brazil}
\author{Luciano R. da Silva}
\affiliation{Federal University of Rio Grande do Norte, Departamento de F\'isica Te\'orica e Experimental, Natal-RN, 59078-900, Brazil.}
\affiliation{National Institute of Science and Technology of Complex Systems, Brazil.}
\author{Constantino Tsallis}
\affiliation{National Institute of Science and Technology of Complex Systems, Brazil.}
\affiliation{Centro Brasileiro de Pesquisas F\'isicas, Rua Xavier Sigaud 150, 22290-180 Rio de Janeiro-RJ, Brazil.}
\affiliation{Santa Fe Institute, 1399 Hyde Park Road, New Mexico 87501, USA}
\affiliation{ Complexity Science Hub Vienna, Josefstaedter Strasse 39, A 1080 Vienna, Austria.}
\date{\today}

\begin{abstract}
Systems which consist of many localized constituents interacting with each other can be represented by complex networks. Consistently, network science has become highly popular in vast fields focusing on natural, artificial and social systems. We numerically analyze the growth of $d$-dimensional geographic networks (characterized by the index $\alpha_G\geq0$; $d = 1, 2, 3, 4$) whose links are  weighted through a predefined random probability distribution, namely $P(w) \propto e^{-|w - w_c|/\tau}$, $w$ being the weight $ (w_c \geq 0; \; \tau > 0)$. In this model, each  site has an evolving degree $k_i$ and a local energy $\varepsilon_i \equiv \sum_{j=1}^{k_i} w_{ij}/2$ ($i = 1, 2, ..., N$) that depend on the weights of the links connected to it. Each newly arriving site links to one of the pre-existing ones through preferential attachment given by the probability $\Pi_{ij}\propto \varepsilon_{i}/d^{\,\alpha_A}_{ij} \;\;(\alpha_A \ge 0)$, where $d_{ij}$ is the Euclidean distance between the sites. Short- and long-range interactions respectively correspond to $\alpha_A/d>1$ and $0\leq \alpha_A/d \leq 1$;  
$\alpha_A/d \to \infty$ corresponds to interactions between close neighbors, and $\alpha_A/d \to 0$ corresponds to infinitely-ranged interactions. The site energy distribution $p(\varepsilon)$ corresponds to the usual degree distribution $p(k)$ as the particular instance $(w_c,\tau)=(1,0)$. We numerically verify that the  corresponding connectivity distribution $p(\varepsilon)$ converges, when $\alpha_A/d\to\infty$, to the weight distribution $P(w)$ for infinitely wide distributions (i.e., $\tau \to \infty, \,\forall w_c$) as well as for $w_c\to0, \, \forall\tau$. Finally, we show that $p(\varepsilon)$ is well approached by the $q$-exponential distribution $e_q^{-\beta_q |\varepsilon - w_c^{\prime}|}$ [$0 \leq w_c^{\prime}(w_c, \alpha_A/d) \leq w_c$] which optimizes the nonadditive entropy $S_q$ under simple constraints; $q$ depends only on $\alpha_A/d$, thus exhibiting universality. 
\end{abstract}

\maketitle

\section{Introduction}

Network science is extremely effective in studying large interacting systems. Within this approach, systems are described by graphs, with nodes (or sites) representing the individual components and edges (or links) representing the interactions between them. Empirical studies show that complex networks are ubiquitous \cite{mason2007graph, proulx2005network, wasserman1994social}. It has  been possible, along such lines, to understand the propagation of information \cite{boccaletti2006complex, lind2007spreading}, classical and quantum internet connections~\cite{tilch2020multilayer, brito2020statistical}, scientific collaborations~\cite{newman2001structure, newman2001scientific}, epidemiology~\cite{danon2011networks, firth2020using},  and human brain~\cite{sporns2005human}. Consistently, in recent decades, mathematical models have emerged to mimic real systems and reproduce their structural properties ~\cite{newman2003structure}. In Euclidean space, geographical networks have also been studied, such as subway systems~\cite{latora2002boston}, neural~\cite{sporns2002network}, internet and transportation networks~\cite{gastner2006spatial}. Also, many real world networks are weighted by assigning real numbers to their edges~\cite{newman2004analysis, allard2017geometric}. For example, the US air transportation network~\cite{cheung2012complex}, where the weights of the edges represent the total number of passengers.

Statistical mechanics is intensively used to study systems with complex geometric and topological properties. In many such systems, the elements exhibit long-range interactions~\cite{campa2014physics}. To handle these cases, a generalization of the Boltzmann–Gibbs (BG) statistical mechanics was proposed in $1988$ ~\cite{tsallis1988possible}, currently referred to as nonextensive statistical mechanics, based on the nonadditive entropies $S_q = k \frac{1 - \sum_i p_i^q}{q - 1}$ with $q\in\Re$. The BG theory is recovered for $q\to1$. This generalized approach has been applied in a large variety of systems, e.g.,  spin-glasses~\cite{pickup2009generalized},  astrophysical plasma~\cite{livadiotis2011invariant}, urban agglomerations~\cite{malacarne2001q}, velocities of collective migrating cells~\cite{lin2020universal}, cold atoms in dissipative optical lattices~\cite{douglas2006tunable, lutz2013beyond}, among others.

The relationship between asymptotically scale-free $d$-dimensional  geographic networks and nonextensive statistical mechanics started being explored in 2005~\cite{soares2005preferential, BritoSilvaTsallis2016, nunes2017role, brito2019scaling, cinardi2020generalised, de2021connecting}, where a preferential attachment index $\alpha_A$ and a growth index $\alpha_G$ were included. These studies showed that geographic networks exhibit three regimes: (a) $0 \leq \alpha_A/d \leq 1$, corresponding to strongly long-range interactions, (b) $1\leq\alpha_A/d \lesssim 5$, corresponding to moderately long-range interactions, and (c) $\alpha_A/d \gtrsim 5$, corresponding to the BG-like regime, i.e., $q\simeq 1$ (short-range interactions).

In a recent study, we have introduced a $d$-dimensional geographical network with weighted links~\cite{de2021connecting} where we use a stretched-exponential distribution $P(w)$ for the weight $w$. We analyzed the distribution of site 'energies' (or costs) $p(\varepsilon)$, where
\begin{equation}
\label{eq:energy_site}
    \varepsilon_i \equiv \sum_{j=1}^{k_i} \frac{w_{ij}}{2}\,,
\end{equation}
$k_i$ being the degree of the $i$-th site; the factor $1/2$ is introduced to avoid double counting between the sites, where only half of the link width is assigned to the site $i$. We verified that $p(\varepsilon)$ is numerically very close to $p_q(\varepsilon)\equiv e_q^{-\beta_q \varepsilon}/Z_q$, where $p_q(\varepsilon)$ generalizes the BG factor, $\beta_q$ playing the role of an inverse temperature and $Z_q$ being the normalization factor; the $q$-exponential function is defined as $e^z_q \equiv [1 + (q- 1)z]^{\frac{1}{1 - q}}$ $(e^z_1 = e^z)$. We also showed that $q$ exhibits a universal behaviour, depending only on the scaled variable $\alpha_A/d$.

Our aim here is to study the energy distribution $p(\varepsilon)$ corresponding to a Laplace-like weight distribution. In particular, we compare $p(\varepsilon)$ with $P(w)$ and with the $q$-exponential form. 

The comparison of distributions is common in diverse scenarios. From algorithms to generate pseudo-random numbers~\cite{thas2010comparing}, through the validity of empirical data with regard to the corresponding theoretical distribution~\cite{young1977proof, peacock1983two} and the identification of images~\cite{swain1991color, yang2002detecting}, to distributions of aquatic organisms~\cite{wiesebron2016comparing}. Interestingly enough, the studies~\cite{yook2001weighted, wang2004weighted} show that the site energy (total weight) distribution  converges (excepting for a logarithmic correction) to the scaling behavior of the connectivity (or degree)  distribution of the network. Empirical studies exhibit some real-world examples of networks where this phenomenon or similar ones have been observed,  e.g., in book borrowing, movie actor collaborations~\cite{wang2004weighted}, worldwide airport connections, scientific collaborations~\cite{barthelemy2005characterization}, and others. In the present work we numerically show that, when ($\tau\to\infty, \forall w_c$) and ($w_c\to0, \forall\tau$),
the energy distribution $p(\varepsilon)$ and the weight distribution $P(w)$ become identical in the regime of short-range interactions ($\alpha_A/d \gg 1$).

The paper is structured as follows. In Sec. II we describe our network model, including the weight distribution and its stochastic implementation. In Sec. III we analyze the numerical energy distribution and compare it with the weight distribution of the links, as well as with the $q$-exponential form emerging from optimizing the nonadditive entropy $S_q$. Finally, in Sec. IV we conclude.

\begin{figure}[!t]
\begin{center}
\includegraphics[scale=.65]{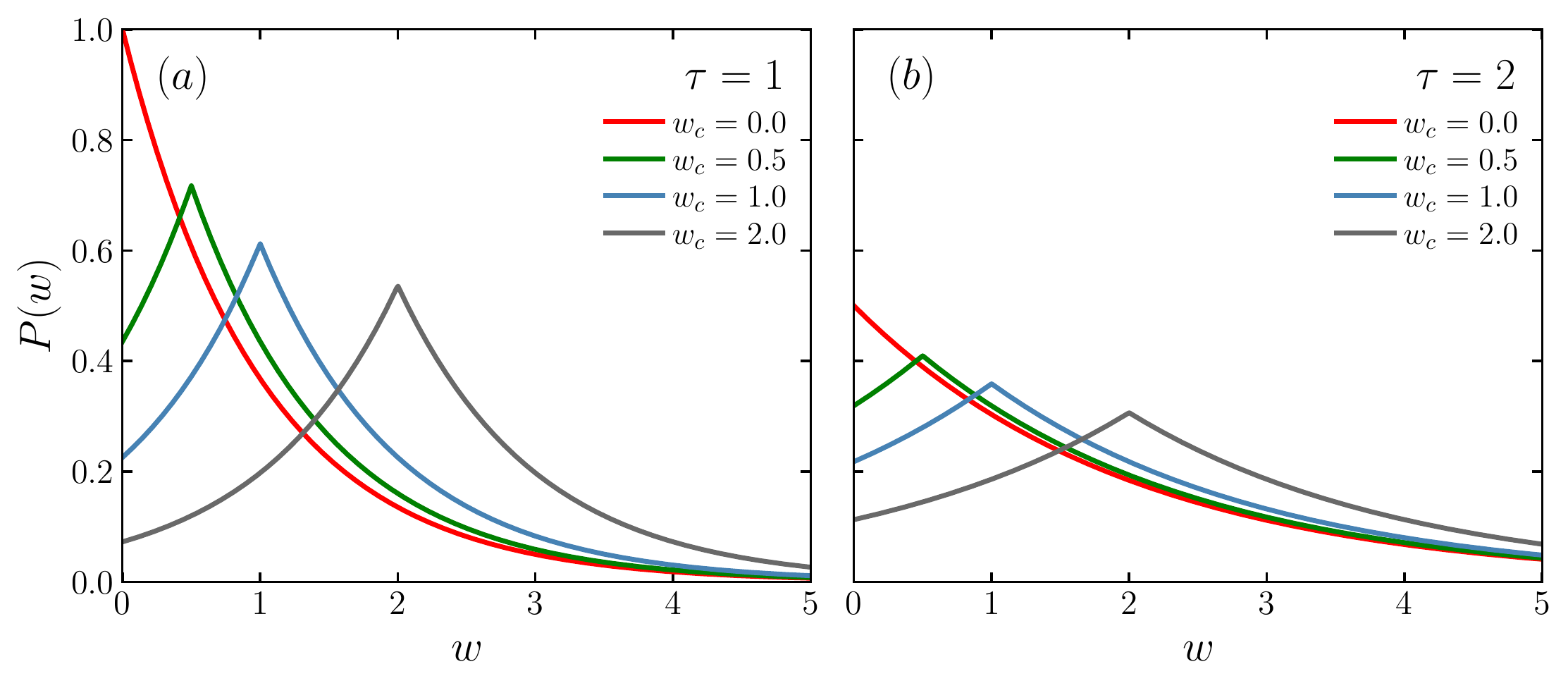}
\end{center}
\caption{
Examples of weight distributions $P(w)\propto e^{- |w - w_c|/\tau}$ used for the network links. This distribution is defined within the range $w\in[0, \infty)$ which characterizes an asymmetric Laplace-like distribution with the free parameters $w_c \geq 0$ and $\tau>0$ that control the location of the  peak and its width (scale parameter) respectively. Its special case $w_c = 0$  corresponds to the exponential distribution. (a) $P(w)$ for $\tau = 1$ and typical values of $w_c = (0, 0.5, 1.0, 2.0)$.  (b) The same as in (a) but with scale parameter $\tau = 2$.
}
\label{pw_laplace}
\end{figure}

\section{The model}

Our network model starts with one site at the origin and then, for each new site added to the network, we randomly sample a position for it according to the isotropic distribution
$p(r) \propto 1/r^{d+ \alpha_G} \;\;\; (\alpha_G > 0$; $d = 1, 2, 3, 4$),
where $r\geq1$ is the Euclidean distance from the center of mass to the new site, and $\alpha_G$ is the growth index. Then, we link every newly arriving site ($j$) to one of the pre-existing ones in the network
according to the following  preferential attachment rule:
\begin{eqnarray}
\Pi_{ij}= \frac{\varepsilon_{i}\,d_{ij}^{-\alpha_A}}{\displaystyle{\sum_{k} \varepsilon_k \, d_{kj}^{-\alpha_A}}}\;\in[0,1]\; \;\;(\alpha_A \ge 0),
\label{attachment}
\end{eqnarray}
where $d_{ij}$ is the Euclidean distance between sites $i$ and $j$ and the attachment index $\alpha_A$ characterizes the range of the interactions; when $\alpha_A\to0$ the system has long-range interactions and the distance loses relevance, whereas for $\alpha_A\to\infty$ the sites have short-range interactions (connections between close neighbors). To each site of the network we associate, through Eq. (\ref{eq:energy_site}), a local \textit{energy} $\varepsilon_i$  which depends on the number of links.

In the present paper, we are using the following  Laplace-like distribution (see Fig. \ref{pw_laplace}) for the weights of the links:
\begin{equation}
\label{eq:weight_links}
    P(w) = {\displaystyle \frac{1}{2\tau - \tau e^{-w_c/\tau}}} e^{-|w - w_c|/\tau} \;\; (w_c \geq 0; \; \tau > 0),
\end{equation}
which satisfies $\int^\infty_0 \mathrm{d} w P(w) = 1$, $\tau$ characterizing the distribution width and $w_c$ being the peak location parameter. The parameter $\tau$ influences the wide of the distribution: when $\tau\to0$ the distribution displays a prominent peak around $w_c$, whereas, in contrast, $\tau\to\infty$ corresponds to an infinitely wide distribution. To numerically get values of the variable $w$ from this distribution we use the inverse transformation method~\cite{devroye1986}:

\begin{equation}
    w  = w_c - \text{sgn}\Bigl[u - A\tau(1 - e^{-w_c/\tau})\Bigl]\tau\ln\biggl\{1 
+ \text{sgn}\Bigl[u - A\tau(1 - e^{-w_c/\tau})\Bigl]\Bigl[1 - (u/A\tau + e^{-w_c/\tau})\Bigl]\biggl\},
\end{equation}
where $A = 1/(2\tau - \tau e^{-w_c/\tau})$ is the normalization constant [see Eq. (\ref{eq:weight_links})], $u$ is a uniform random variable between $[0, 1]$ and $\text{sgn}(\theta)$ denotes the sign function:
\begin{equation*}
    \text{sgn}(\theta) = \begin{cases}
-1, &\text{if}\; \theta < 0,\\
0, &\text{if}\; \theta = 0, \\
1, &\mbox{if}\; \theta > 0.
\end{cases}
\end{equation*}

\begin{figure}[!t]
\begin{center}
\includegraphics[scale=.6]{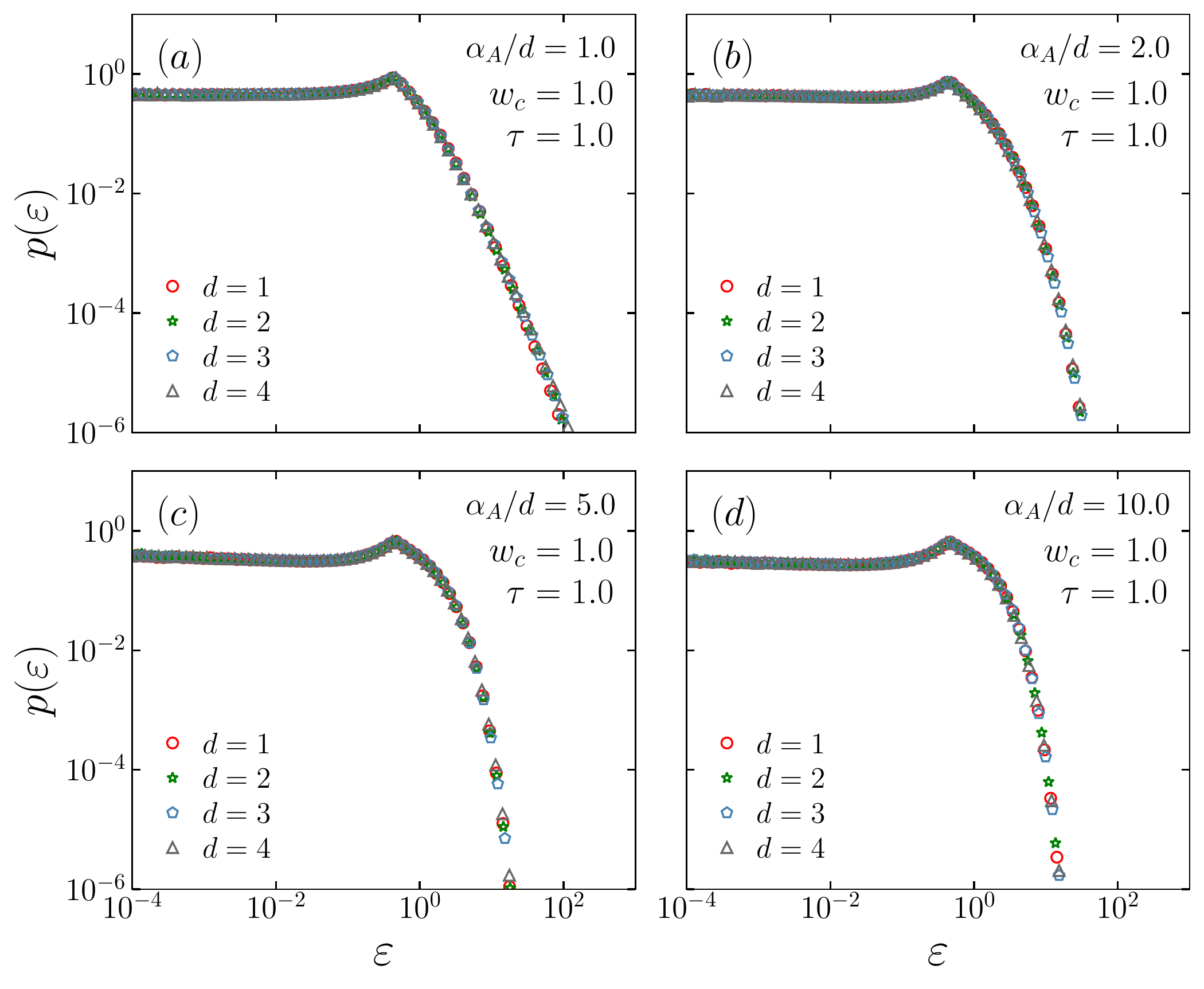}
\end{center}
\caption{Local \textit{energy} distribution $p(\varepsilon)$ for typical values of $d=1,2,3,4$ with $(a)$ $\alpha_A/d = 1$, $(b)$ $\alpha_A/d = 2$, $(c)$ $\alpha_A/d = 5$ and $(d)$ $\alpha_A/d = 10$ by fixing ($\alpha_G, \; \tau, \; w_c$) = ($1, \; 1, \; 1$). We verify that, for all values of $d$, the curves of $p(\varepsilon)$ remain invariant, i.e., the dimensionality does not influence the energy distribution. For simplicity, we set $w_c=\tau=1$, but the results remain independent from this choice. The simulations are averaged over $10^3$ realisations for $N = 10^5$ sites.}
\label{pde_aA_d}
\end{figure}

\section{SIMULATIONS and RESULTS}
We now focus on the energy distribution $p(\varepsilon)$ and compare it with the weight distribution of the links $P(w)$. The weight distribution has two free parameters, namely  $\tau$ and $w_c$,  while our model has three parameters, namely  $\alpha_A$, $\alpha_G$ and $d$, in addition to the weight distribution itself. We fix these parameters and numerically determine $p(\varepsilon)$ by using a large number $N$ of sites (typically $N=10^5$) and performing a large number of realizations (typically $10^3$). 

We verify that the energy distribution is completely independent from $\alpha_G$ in all situations, as already discussed in earlier publications~\cite{soares2005preferential, BritoSilvaTsallis2016, nunes2017role, brito2019scaling, de2021connecting}. Consequently, we fix it to be $\alpha_G = 1$ in all our simulations. Also, we observe that $p(\varepsilon)$ remains invariant when we fix $\alpha_A/d$ and we modify the values of $d=1,2,3,4$, as shown in Fig.~\ref{pde_aA_d}. This implies an universality property, namely that the energy distribution depends on the ratio $\alpha_A/d$ and does not depend independently on $\alpha_A$ and on $d$~\cite{BritoSilvaTsallis2016, nunes2017role, brito2019scaling, de2021connecting, cinardi2020generalised}. In consequence, we present our results by simply running $d=2$.

\begin{figure}[!t]
\begin{center}
\includegraphics[scale=.55]{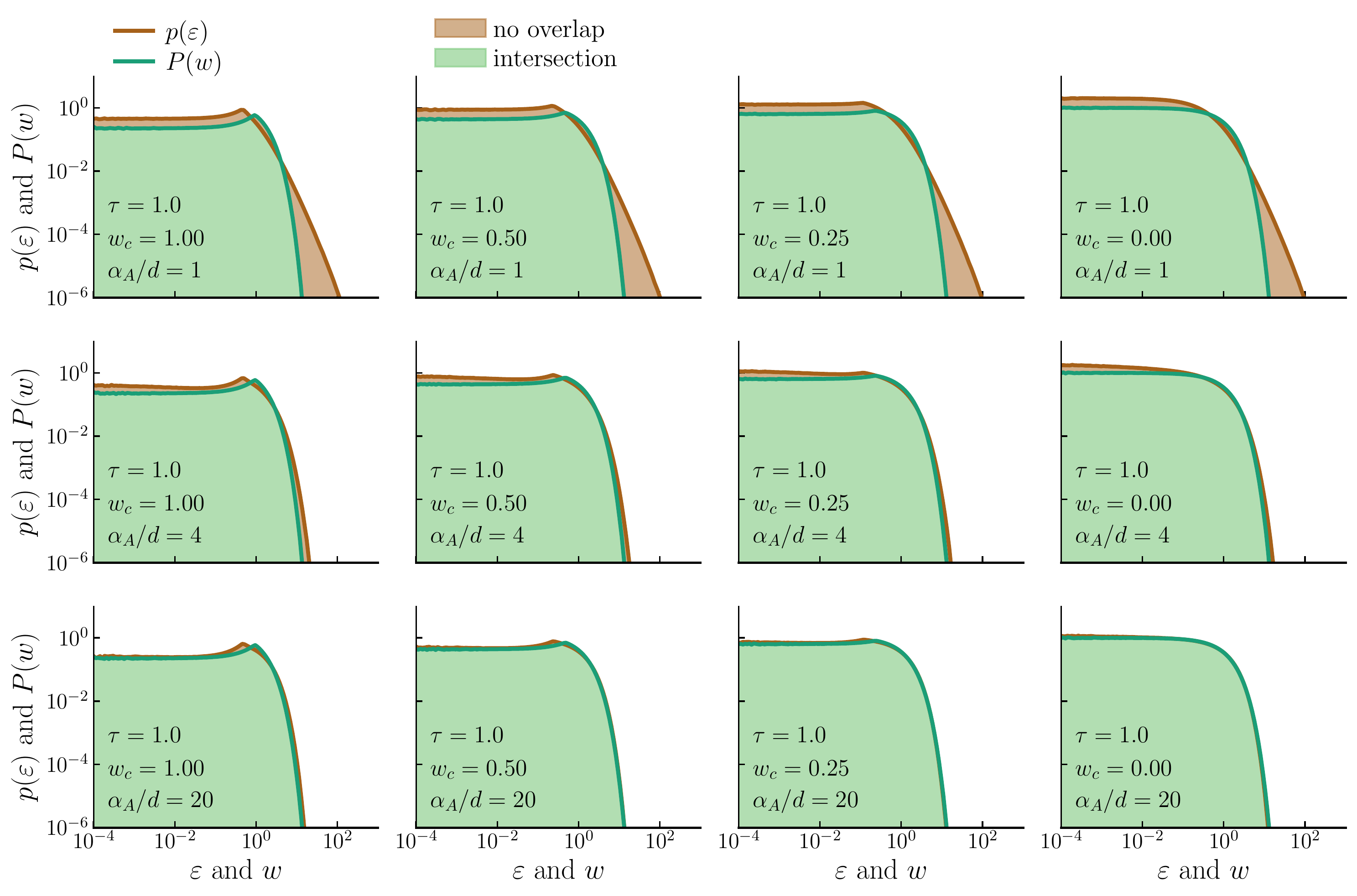}
\end{center}
\caption{Comparison between the distribution $P(w)$ of the associated link weights and the energy distribution $p(\varepsilon)$ of the network. The green and brown continuous lines respectively represent $P(w)$, given by Eq. (\ref{eq:weight_links}), and $p(\varepsilon)$. The green-shaded area is the region of intersection and the brown-shaded area denotes the region where the weight and energy distributions do not intersect. We analyzed a large amount of typical cases with $w_c=$ ($1,\; 0.50,\; 0.25,\; 0$) and $\alpha_A/d=$ ($1,\; 4,\; 20$), with $\tau = 1$. For large values of $\alpha_A/d$ the results show that the distribution curves are similar in the tail with different peaks where the maximum value of the energy distribution is shifted to the left; if $w_c\to0$ the distributions become identical. The simulations are averaged over $10^3$ realisations for $N = 10^5$ sites.}
\label{inter-tau-1}
\end{figure}

We choose typical values for the parameter $\tau = 1$ and $10$, and for the location parameter $w_c = 0$, $0.25$, $0.5$, $1$ and $2$, and generate networks for fixed values of the attachment parameter $\alpha_A$. We compare the weight and the energy distributions through their histograms by using a function from the Python Library Numpy~\cite{harris2020array}. This function has some input parameters (array, bins, density, etc.) and returns the probability density function at each bin (with $n$ bins in total, $b_1, \dots, b_{n}$) where the integral over the entire interval equals unity.

\begin{figure}[!t]
\begin{center}
\includegraphics[scale=.55]{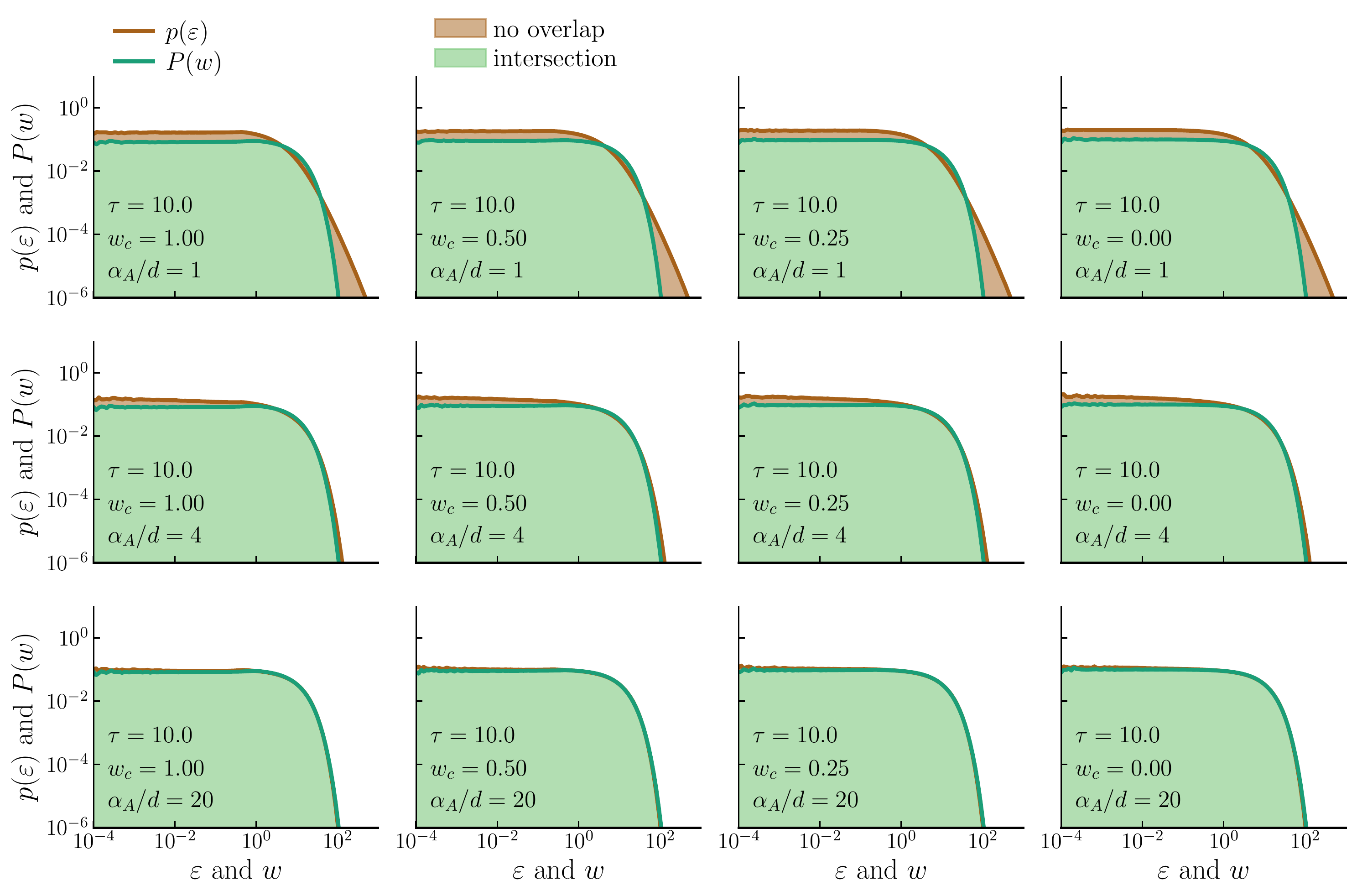}
\end{center}
\caption{The same as in Fig. \ref{inter-tau-1}
but with $\tau =10$.
}
\label{inter-tau-10}
\end{figure}

There are many methods in the literature for comparing histograms. We use here two of them, namely Histogram Intersection~\cite{swain1991color} and Q–Q (quantile-quantile) plot, to compare $p(\varepsilon)$ and $P(w)$. The histogram intersection method is a technique used for image indexing and comparison, where the image colors are discretized 
by a histogram and compared to a original figure. Given two histograms, $H_w$ and $H_{\varepsilon}$, containing the same  number $n$ of bins, Swain and Ballard~\cite{swain1991color} defined the histogram intersection as the sum of the minima for all histogram bins:

\begin{eqnarray}
\label{eq:inter}
H_w \cap H_{\varepsilon} = \sum_{i=1}^{n} \mbox{min}\left[h_w(i), h_{\varepsilon}(i)\right],
\label{intersection}
\end{eqnarray}
the range of this calculus goes from $0$ to $1$, where $0$ indicates that the histograms do not intersect at all and $1$ means that the histograms are identical.

\begin{figure}[!t]
\begin{center}
\includegraphics[scale=.65]{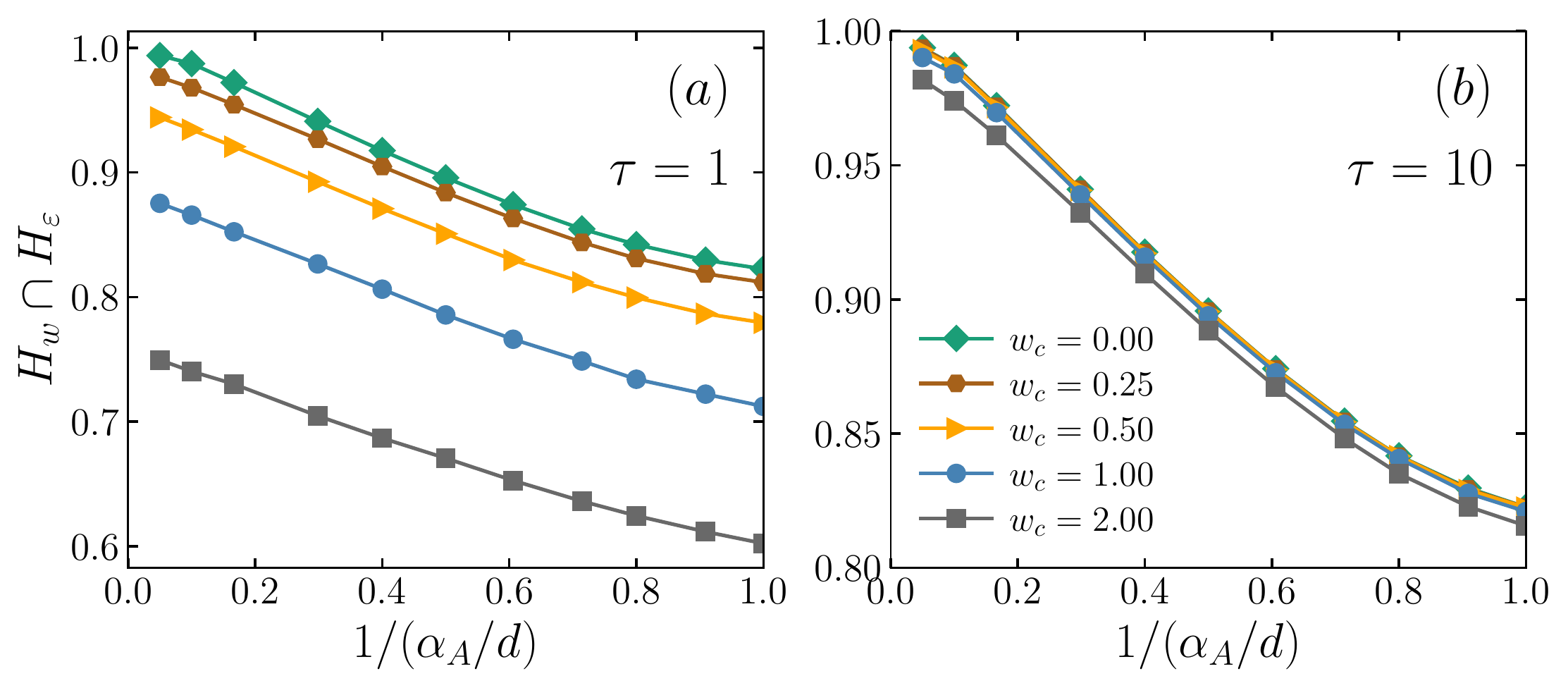}
\end{center}
\caption{ Intersection between the histogram $H_{\varepsilon}$ of site energies  and the histogram $H_w$ of the link weights. Every point was calculated using Eq. (\ref{intersection})  for (a) $\tau = 1$ and (b) $\tau = 2$, and typical values of $w_c = (0, 0.25, 0.5, 1, 2)$. As we can see in (a), all the curves attains the maximum value at $\alpha_A/d\to\infty$, but only for $w_c \sim 0$ the intersection between histograms approaches $1$, showing that the distributions $p(\varepsilon)$ and $P(w)$ are practically identical when $\tau = 1, \; w_c\sim0$. We can observe in (b) that the curves exhibit a maximum close to $1$ regardless of the location parameter  $w_c$. This result is a consequence from the fact that $P(w)$ becomes an infinitely wide distribution, i.e.,  when $\tau\to\infty$, and, simultaneously,  $\alpha_A/d\to\infty$, i.e., the network exhibits connectivity only between sites that are very close.}
\label{inter-wc}
\end{figure}

In Figs.~\ref{inter-tau-1} and \ref{inter-tau-10} we show the energy and weight distributions for $\tau = 1$ and $\tau = 10$ respectively. We observe that when $\alpha_A/d\sim1$ the histograms are sensibly different, whereas when $\alpha_A/d$ increases they become quite similar. In the limit $\alpha_A/d\to\infty$, the distributions are practically identical. This behaviour is shown in Fig.~\ref{inter-wc}. This result is more pronounced when $w_c\to0$.

\begin{figure*}[!t]
\begin{center}
\includegraphics[scale=.59]{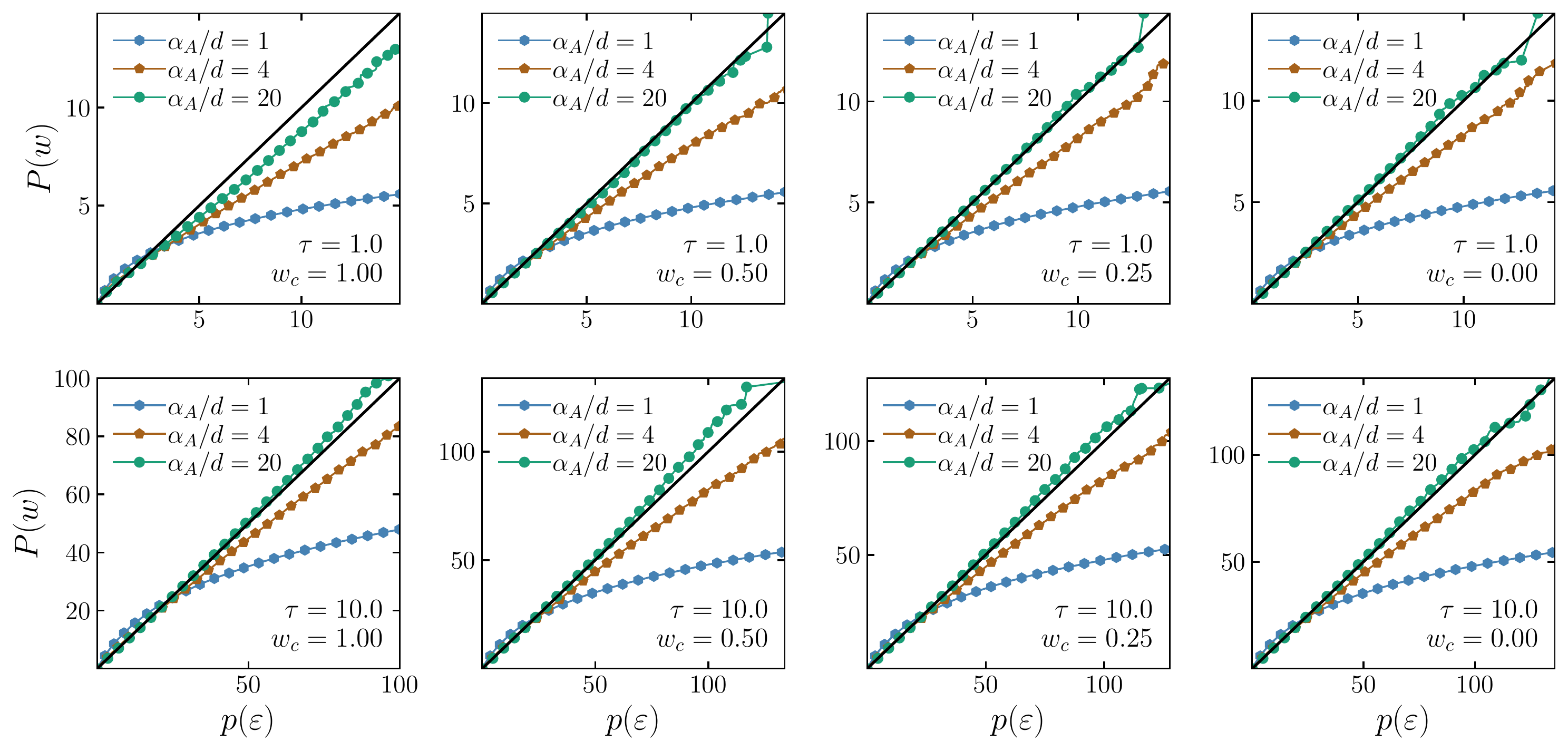}
\end{center}
\caption{Q-Q plot for $\tau = 1$ (top) and $\tau = 10$ (bottom), and
typical values of $\alpha_A/d = 1$ (blue diamond), $4$ (brown hexagon), $20$ (green circle). In all cases the black continuous line corresponds to the bisector. The simulations were averaged over $800$ realisations for $N = 10^5$ sites.}
\label{qqplot}
\end{figure*}

Alternatively, we can also recover the same results through the quantile probability plot (Q–Q plot), which is a graphical method for comparing two probability distributions by plotting the quantile of the first distribution against the same quantile of the second one. If $F(x)$ is a distribution function, the $p^{th}$ quantile of F is defined as~\cite{serfling2009approximation}:
\begin{equation}
    \xi_p \equiv \inf \{x: F(x) \leq p \} \;\; (0<p<1).
\end{equation}

If the two distributions are very close, the points in the Q–Q plot will fall approximately along a straight line, namely the bisector. Otherwise, if the  points form a nonlinear curve, then the distributions are different. Figure \ref{qqplot} presents a Q-Q plot for the energy of the sites against the weights of the links. We observe that the quantile probability plot accentuates the comparative structure of the tails of the variables  $\varepsilon$ and $w$. When $\alpha_A/d = 1$ the discrepancy between the distributions is clearly evident, where the weight distribution has a considerably shorter tail than the energy distribution.

\begin{table}
\centering
\caption{Values used for $\beta_{q_0}$ and $\beta_{q_\infty}$ in Eq. (\ref{eq:beta}). The parameters $\beta_{q_0}$ and $\beta_{q_\infty}$ correspond to $\alpha_A/d = 1$ and $\alpha_A/d = 20$ respectively. These results are for $N = 10^5$ sites averaged over $10^3$ realizations.} 
\label{table:beta}
\begin{tabular}{p{10em} p{10em} p{5em}} 
 \hline
$w_c$ & $\beta_{q_0}$ & $\beta_{q_\infty}$\\ [0.5ex] 
 \hline
 $0.00$ & $3.33$ & $1.14$\\ 
$0.25$ & $2.80$ & $1.04$\\
$0.50$ & $2.55$ & $1.00$\\
 \hline
\end{tabular}
\end{table}

\begin{figure*}[!t]
\begin{center}
\includegraphics[scale=.56]{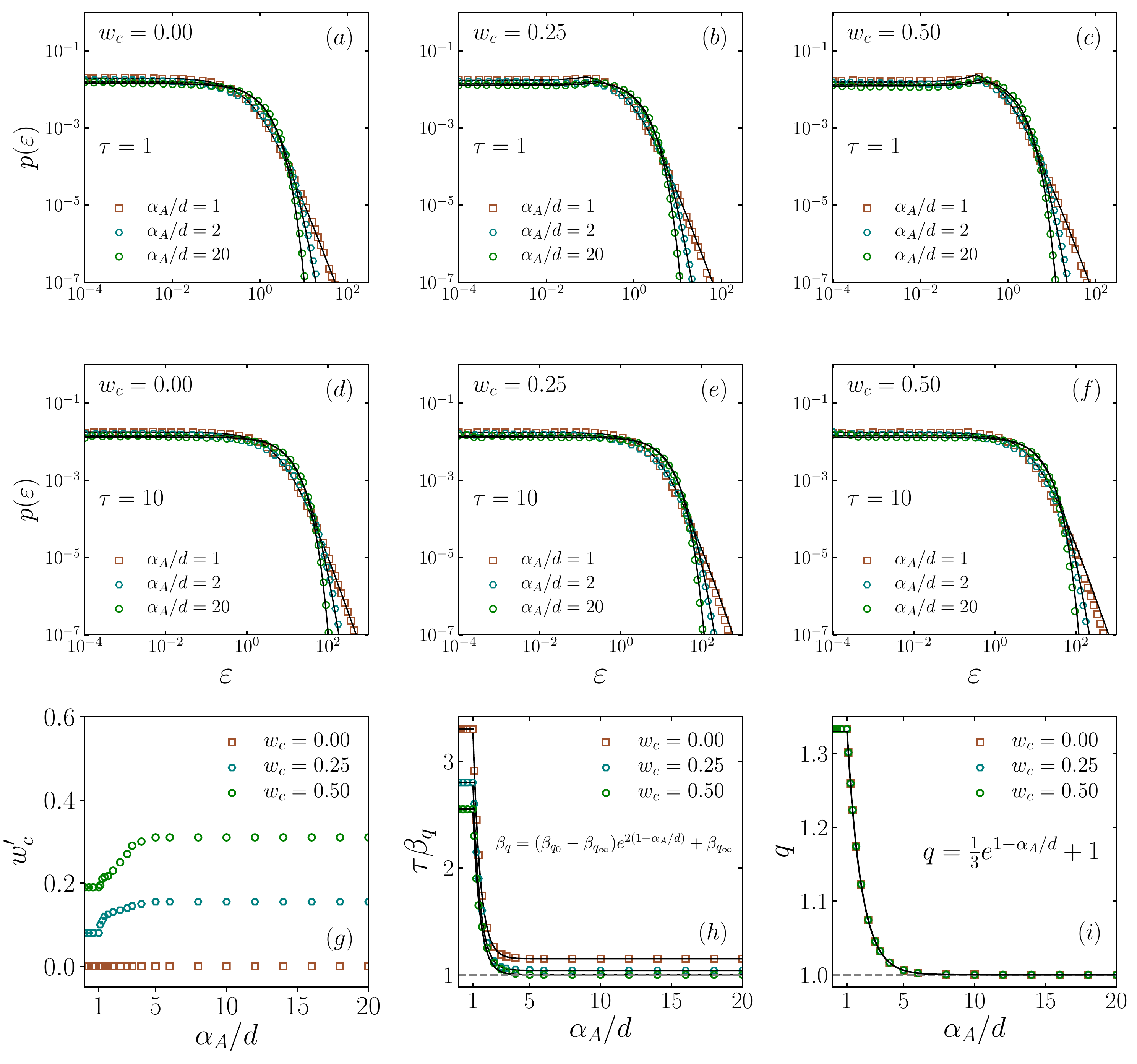}
\end{center}
\caption{\textit{Top:} \textit{Energy} distribution for   $w_c = 0.00$ $(a)$,   $w_c = 0.25$ $(b)$ and  $w_c = 0.50$ $(c)$, and $\alpha_A/d = 1, 2, 20$, with $\tau = 1$. The plots $(d)$-$(f)$ are the same as in $(a)$-$(c)$ but with $\tau = 10$. The black continuous lines correspond to Eq. (\ref{fitting:pe}) with $q$ and $\beta_q$ given by Eqs. (\ref{eq:q}) and (\ref{eq:beta}) respectively. \textit{Bottom:} $(d)$ $w_c^{\prime}$ as function of $\alpha_A/d$ for typical values of $w_c$; $w_c^{\prime}$ is constant for $0 \leq \alpha_A/d \leq 1$, increases for $1< \alpha_A/d \lesssim 5$,  and attains a constant limit for $\alpha_A/d\to\infty$. $(e)$ $\tau\beta_q$ as function of $\alpha_A/d$ for typical values of $w_c$, the black continuous lines correspond to Eq. (\ref{eq:beta}) with $\beta_{q_0}$ and $\beta_{q_\infty}$ listed in Table~\ref{table:beta}. $(f)$ $q$ as function of $\alpha_A/d$ and $w_c =$ $0.0,\; 0.25,\; 0.5$; we verify that $q=4/3$, $\forall \, 0 \le \alpha_A/d < 1$, and is given by the expression which is indicated therein for $\alpha_A/d \ge 1$.  These results remain as they stand for all values of $\tau$.}
\label{fitting}
\end{figure*}

Let us now compare $p(\varepsilon)$ with the following $q$-exponential form:

\begin{equation}
   p_q(\varepsilon) = \frac{e_q^{-\beta_q |\varepsilon - w_c^{\prime}|}}{Z_q},
   \label{fitting:pe}
\end{equation}
where $w_c^{\prime}\in[0, w_c]$ is a location parameter playing the role of a chemical potential in $p_q(\varepsilon)$, $q$ is the entropic index, $\beta_q$ playing the role of an inverse temperature, and $Z_q$ is the normalization factor (see Fig.~\ref{fitting}a-c); we remind that the weight distribution analysed in the present paper differs from that analysed in \cite{de2021connecting}. The particular case $w_c = 0$ corresponds to an exponential distribution and we recover the results obtained in our previous study~\cite{de2021connecting}, where we used the stretched-exponential distribution for the link weight $P(w)\propto e^{-(w/w_0)^{\eta}}$ for $\eta = 1$. Once again, this result exhibits the emergence of an interesting correspondence between a geometrical random network problem and a particular case within nonextensive statistical mechanical systems.

In Fig.~\ref{fitting}(d-f) we exhibit the variables $q$, $\beta_q$ and $w_c^{\prime}$ as functions of the ratio $\alpha_A/d$. We numerically show that $q$ only depends on this ratio; $\beta_q$ also depends on $w_c$. Interestingly enough, both variables are given by the same equations indicated in \cite{de2021connecting}, i. e.,

\begin{eqnarray}
q =
\begin{cases}
\frac{4}{3} &\mbox{if}\; 0\leq \frac{\alpha_A}{d} \leq 1\\
\frac{1}{3}\,e^{1-\alpha_A/d} + 1 &\mbox{if}\; \frac{\alpha_A}{d} > 1
\end{cases}
\label{eq:q}
\end{eqnarray}
and
\begin{eqnarray}
\tau\beta_q =
\begin{cases}
\displaystyle\beta_{q_0} &\mbox{if } 0 \leq \frac{\alpha_A}{d} \leq 1\\
\displaystyle (\beta_{q_0} - \beta_{q_\infty})\,e^{2(1-\alpha_A/d)} +\beta_{q_\infty}&\mbox{if } \frac{\alpha_A}{d} > 1,
\end{cases}
\label{eq:beta}
\end{eqnarray}
where the variables $\beta_{q_0}$ and $\beta_{q_\infty}$ are listed in Table~\ref{table:beta}. This fact was not necessarily expected a priori, and it probably appears because, in both cases, we use $q$-exponentials to approach our results. In all cases we observe the existence of three regimes, consistent with \cite{BritoSilvaTsallis2016, brito2019scaling, de2021connecting}. For $0 \leq \alpha_A/d \leq 1$, $q$, $\beta_q$ and $w_c^{\prime}$ are constants, the system presenting very-long-range interactions. For $1< \alpha_A/d \lesssim 5$, the system exhibits moderately-long-range interactions, where $q$ and $\beta_q$ monotonically decrease with $\alpha_A/d$, whereas $w_c^{\prime}$ increases. In the limit of $\alpha_A/d\to\infty$, the parameters tend to constant values and this regime corresponds to short-range interactions ($q\to1$).

\section{Conclusions}

Many real systems contain sites  whose  importance (here denoted as \textit{energy}) is generated by  interactions with the other sites through weighted links. This is the kind of situation that our network model mimics. Our simulations yield energy distributions which depend on $\alpha_A$ and on $d$ through the ratio $\alpha_A/d$. Indeed, the results collapse into a single curve for any spatial dimension. In addition to that, $p(\varepsilon)$ does not depend on $\alpha_G$. We have also shown that the energy distribution $p(\varepsilon)$ is well approached by the form $e_q^{-\beta_q |\varepsilon - w_c^{\prime}|}$, where $e_q^x$ is the $q$-exponential function emerging within $q$-statistics~\cite{tsallis1988possible}, $\beta_q$ playing the role of an inverse temperature. It is quite remarkable that the $(\alpha_A/d)$-dependence of $q$ precisely coincides with that obtained in \cite{de2021connecting} for a different distribution $P(w)$. This reinforces the strength of an universality conjecture concerning the entropic index $q$. In particular, Eqs. (\ref{eq:q}) and (\ref{eq:beta}) appear to have a quite generic validity.

The distribution of the link weights decays exponentially. It is therefore interesting to observe that, in sensible contrast, the cooperative effect of all of the links  results in a power-law-decaying distribution for the energies. This is somewhat reminiscent of the contrast between the Einstein \cite{ashcroft1976solid} and the Debye \cite{debye1912theorie} models for solids. Indeed, the low-temperature specific heat of independent quantum harmonic oscillators (first approach for optical phonons) displays an exponential behavior while the low-temperature specific heat of coupled harmonic oscillators (first approach for acoustic phonons) follows a power law. This type of cooperative phenomenon in the network has already been observed in our previous work \cite{de2021connecting}. Consequently, it is natural that we compare this result with the $q$-exponential function, which is a power law  with an asymptotic slope equal to $-1/(q - 1)$.

We have also shown that, in the regime of short-range interactions ( $\alpha_A/d\to\infty$), the energy distribution coincides with the weight distribution for a parameter class ($\tau\to\infty$, $w_c\to0$). Based on the intersection of the histograms of those two distributions, and also on the quantile-quantile plot, we have provided strong numerical evidence that this is indeed so for the distribution given in Eq. (\ref{eq:weight_links}). It might well be that this result is more general than here verified, i.e., the same result might be true for  other distributions $P(w)$. This type of theory applies to systems similar to mobile phone call records, where the weights of the links $w_{ij}$ are the total duration of calls (in seconds) between the users $i$ and $j$. For this type of network, the weight and the energy (or strength) distributions are similar and the widths of the links are correlated with the energies of the sites \cite{onnela2007analysis}.

Let us conclude by a rather general comment. Various examples are known where Boltzmann-Gibbs statistical mechanical systems are isomorphic to random geometrical problems. Such is the case of the Kasteleyn-Fortuin theorem~\cite{kasteleyn1969phase}, 
where the $q_{\text{Potts}}\to1$ limit of the $q_{Potts}$-state Potts ferromagnet corresponds to bond percolation. That is also the case of the de Gennes isomorphism \cite{de1972exponents}, where the $n\to 0$ limit of the $n$-vector ferromagnet corresponds to self-avoiding random walk, cornerstone of polymer physics. 
Our present numerical results suggest that analogous connections appear to exist between nonextensive statistical mechanical systems and (asymptotically) scale-free random networks. 
\section*{Acknowledgements}
We thank the High Performance Computing Center (NPAD/Universidade Federal do Rio Grande do Norte) for providing computational resources. Partial financial support from CNPq, Capes and Faperj (Brazilian agencies) is acknowledged as well.

\end{document}